\documentclass[showpacs,showkeys,nopreprintnumbers,prl,twocolumn]{revtex4-1}%
\usepackage{amssymb}
\usepackage{amsfonts}
\usepackage{amsmath}
\usepackage{graphicx}%
\providecommand{\U}[1]{\protect\rule{.1in}{.1in}}

\begin{document}
\preprint{HEP/123-qed}
\title[ ]{In-plane optical spectra of Y$_{1-x}$Ca$_{x}$Ba$_{2}$Cu$_{3}$O$_{7-\delta}$:
Overdoping and disorder effect on residual conductivity}
\author{E. Uykur}
\affiliation{Department of Physics, Graduate School of Science, Osaka University, Osaka
560-0043, Japan}
\author{K. Tanaka}
\affiliation{Department of Physics, Graduate School of Science, Osaka University, Osaka
560-0043, Japan}
\author{T. Masui}
\affiliation{Department of Physics, Graduate School of Science, Osaka University, Osaka
560-0043, Japan}
\author{S. Miyasaka}
\affiliation{Department of Physics, Graduate School of Science, Osaka University, Osaka
560-0043, Japan}
\author{S. Tajima}
\affiliation{Department of Physics, Graduate School of Science, Osaka University, Osaka
560-0043, Japan}
\keywords{(Y,Ca)Ba$_{2}$Cu$_{3}$O$_{7-\delta}$, far-infrared, in-plane optical spectra,
chain conductivity}
\pacs{74.25.Gz, 74.72.-h}

\begin{abstract}
We measured the temperature dependence of the in-plane polarized reflectivity
spectra of twin-free Y$_{1-x}$Ca$_{x}$Ba$_{2}$Cu$_{3}$O$_{7-\delta}$ single
crystals with different Ca-concentrations ($x=0,0.11\ $and $0.16$) from
optimally doped to heavily overdoped region. Low energy optical conductivity
spectra showed a Drude-like residual conductivity at temperatures far below
the superconducting transition temperature, which indicates the presence of
unpaired-normal carriers in the superconducting state. Comparing the spectra
at a fixed Ca-content or at a fixed doping level, we have revealed that the
carrier overdoping increases unpaired carriers in addition to those induced by
the Ca-disorder. We also found the superconducting behavior of the
one-dimensional CuO chains for the Ca-free samples.

\end{abstract}
\volumeyear{year}
\volumenumber{number}
\issuenumber{number}
\eid{identifier}
\startpage{1}
\endpage{2}
\maketitle

\section{Introduction}

The overdoped regime of cuprate superconductors has not been intensively
investigated, compared to the well studied pseudogap phase in the underdoped
regime \cite{pseudogap state}, although many unresolved problems remain in the
former. One of the reasons for this situation is that it is believed that in
the overdoped regime, the system is approaching a conventional Fermi liquid
state and thus nothing peculiar happens. The other is a technical reason that
only limited materials such as Tl-based cuprates can achieve the overdoped state.

Among some anomalies in the overdoped regime, we focus here on the low energy
optical response. When the system goes into the superconducting state with
\textit{d}-wave symmetry, the optical conductivity is suppressed below the gap
energy ($2\Delta$) and gradually decreases towards zero at $\omega=0$.
However, in many cuprates, a substantial amount of conductivity remains finite
at $\hbar\omega\ll2\Delta$, forming a Drude-like spectrum even at $T\approx0$.
This residual conductivity suggests the presence of unpaired normal carriers
in the superconducting state. Impurity pair-breaking could be one of the
sources for this residual conductivity, since Zn-doped cuprates exhibit a
similar low frequency response \cite{ZnYBCO2}. The problem is that this
anomaly can be observed even in impurity-free samples \cite{RC in all HTSC}.
Its origin is unknown.

The increase of the residual conductivity with carrier overdoping was first
reported in the \textit{c}-axis spectra of YBa$_{2}$Cu$_{3}$O$_{7-\delta}$
(Y-123) \cite{rcYBCO1, rcYBCO2} and Y$_{1-x}$Ca$_{x}$Ba$_{2}$Cu$_{3}%
$O$_{7-\delta}$ (Y/Ca-123) \cite{rcCa-YBCO}, followed by the study of the
in-plane spectra of Tl$_{2}$Ba$_{2}$CuO$_{6+\delta}$ \cite{rcTL2201}. It seems
to be related to overdoping, but not conclusive because disorder-induced
pair-breaking cannot be ignored in most cases. For example, Ca-substitution
for Y in YBa$_{2}$Cu$_{3}$O$_{7-\delta}$ lowers the maximum \textit{T}$_{c}$
value presumably because it introduces disorders into the CuO$_{2}$-planes,
while it also introduces carrier-holes into the system. Therefore, we need to
distinguish the effects of carrier-overdoping and Ca-disorder in order to
discuss the origin of residual conductivity.

So far, there have been few studies for the in-plane optical spectra of
overdoped cuprates except for Tl$_{2}$Ba$_{2}$CuO$_{6+\delta}$ (Tl2201) and
Bi$_{2}$Sr$_{2}$CaCu$_{2}$O$_{8+\delta}$ (Bi2212)\cite{rcTL2201, rcBi2212}. In
the present work, we prepared detwinned single crystals of Y$_{1-x}$Ca$_{x}%
$Ba$_{2}$Cu$_{3}$O$_{7-\delta}$ for various Ca- and oxygen-contents. In order
to separate the effects of overdoping and Ca-disorder, we studied the spectral
change with Ca-content by fixing the doping level, as well as the change with
doping by fixing the Ca-content. These comparisons have revealed that
overdoping definitely increases the unpaired carriers. While Ca-substitution
also enhances residual conductivity, its effect is weaker than the
Zn-substitution effect.

Moreover, for highly oxygenated Y-123, we observed a clear conductivity
suppression due to the superconducting gap opening in the Cu-O chain spectra.
It seems that the decrease of oxygen deficiency leads to high conductivity in
the chains and thus to a superconducting response.

\section{Experiments and Sample Characterization}

Y$_{1-x}$Ca$_{x}$Ba$_{2}$Cu$_{3}$O$_{7-\delta}$ single crystals (for $x=0,$
$0.11$ and $0.16$) were grown by using a pulling technique explained elsewhere
\cite{pulling technique}. As-grown crystals were cut into pieces with
\textit{ab}-plane surfaces that were not smaller than $4\times4$ mm$^{2}$.
Then the pieces were detwinned under uniaxial pressure $\left(  \sim8\text{
kg/mm}^{2}\right)  $. For reflectivity measurements, the sample surfaces were
mechanically polished by using Al$_{2}$O$_{3}$ powder gradually as fine as
$0.3$ $\mu$m. To obtain optimally doped crystals, samples were annealed at
$500$ and $575$ $^{\circ}$C for $x=0$ and $x=0.11$, respectively, in an oxygen
atmosphere for about 1 month. For the overdoped region, all samples were
annealed at $350$ $^{\circ}$C under oxygen flow for about 2 months. The
superconducting transition temperatures were determined by the dc
susceptibility measurements as follows: $93.5$ and $87$ K for $x=0$ and
$x=0.11$, respectively, in the optimally doped regime, while in the overdoped
region, \textit{T}$_{c}$ values decreased down to $90$ K for $x=0$, $75$ K for
$x=0.11$, and $70$ K for $x=0.16$. The doping levels (\textit{p}) were
determined from \textit{T}$_{c}$, assuming the empirical formula between
\textit{p} and \textit{T}$_{c}$ \cite{p-Tc}. DC resistivity was measured by a
standard four probe method.

The temperature dependent reflectivity measurements were performed with a
Bruker 80v Fourier transform infrared (FTIR) spectrometer from $\sim70$ to
$\sim20000$ cm$^{-1}$ with \textit{E}//\textit{a} and \textit{E}//\textit{b}
polarizations at various temperatures from $\sim10$ to $\sim300$ K. The sample
and the reference mirror (Au for far-infrared and middle infrared regions, Ag
for near infrared and visible regions) were placed into a He-flow cryostat.
Their spectra were compared successively at each measured temperature by
checking their positions with a He-Ne laser. The optical conductivity spectra
were obtained from the measured reflectivity spectra by using the
Kramers-Kronig (K-K) transformation with the Hagen-Rubens extrapolation in the
low-energy region. For the analysis, we used room-temperature reflectivity
spectra in the higher-energy region up to $40$ eV, which were measured with
the use of synchrotron radiation at UV-SOR, Institute for Molecular Science
(Okazaki). We also used $\omega^{-4}$ extrapolation above $40$ eV.%

\begin{figure}[ptb]%
\centering
\includegraphics[
height=2.3964in,
width=3.3918in
]%
{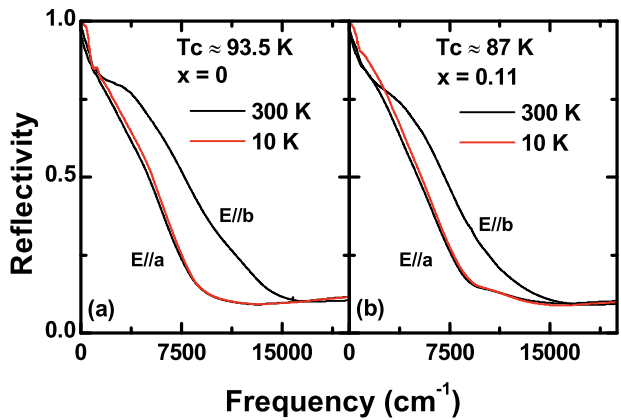}%
\caption{The polarized in-plane reflectivity spectra in the optimally doped
samples with (a) $x=0$ and (b) $x=0.11$.}%
\label{detwinn}%
\end{figure}

Figures \ref{detwinn}(a) and (b) show the in-plane polarized reflectivity
spectra of the optimally doped crystals with $x=0$ and $0.11$ in a wide
frequency range. The $10$ K spectra are also given for \textit{E}//\textit{a}
to demonstrate that the appreciable temperature dependence is observed only in
the low energy region. A clear \textit{a}-\textit{b} anisotropy can be seen
both samples, which guarantees that our detwinning treatment was successful.
The \textit{b}-axis reflectivity edge for the Ca-substituted sample is located
at lower energy compared to the Ca-free sample. This shift can be explained by
the decrease of the carrier concentration in the chains, because in order to
keep a constant doping level, we reduced the oxygen concentration in the
chains for the Ca-substituted samples \cite{Fisher}. The estimated $\delta
$-values were $0.12$ for $x=0$ and $0.257$ for $x=0.11$, respectively.

\section{Results}

\subsection{Overdoping effect on \textit{a}-axis residual conductivity}

\begin{figure}[ptb]%
\centering
\includegraphics[
height=2.3964in,
width=3.3918in
]%
{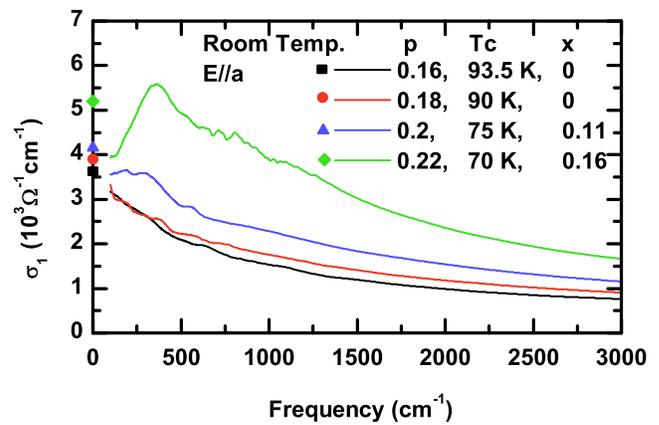}%
\caption{Room temperature \textit{a}-axis optical conductivity from $p=0.16$
to $p=0.22$. Solid symbols indicate the dc value for each sample.}%
\label{comparison}%
\end{figure}

In the overdoped region, the samples\ with $x=0$, $0.11$, and $0.16$ have been
studied. The doping levels (\textit{p}) were determined as $0.18$, $0.2$, and
$0.22$ for these samples, respectively. Figure \ref{comparison} shows the
comparison of the \textit{a}-axis $\sigma_{1}(\omega)$ spectra at room
temperature, together with the optimal doping spectrum for $x=0$. The increase
of the optical conductivity demonstrates a successful increase in carrier
concentration. For the Ca-substituted samples, a peak-like structure evolves
in the low-energy region ($\sim300$ cm$^{-1}$) in the normal state, which
becomes more pronounced with increasing Ca-concentration. With decreasing
temperature, this peak structure merges into the Drude-like conductivity. This
feature is not only dependent of the Ca-content but also of the oxygen
content. The origin of this Ca-induced peak is unknown at the moment.%

\begin{figure}[ptb]%
\centering
\includegraphics[
height=5.585in,
width=3.3927in
]%
{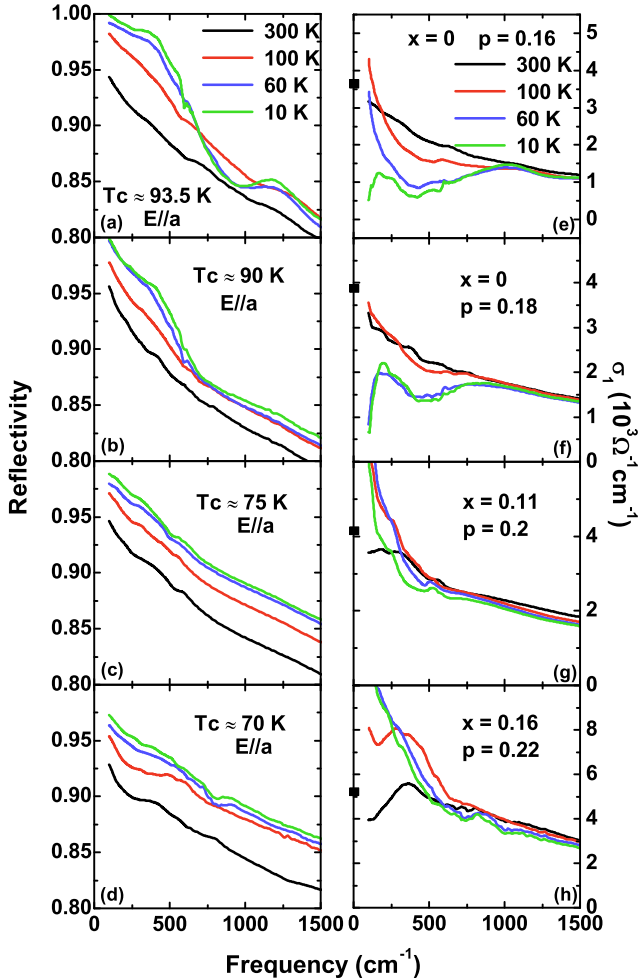}%
\caption{Temperature dependent \textit{a}-axis reflectivity spectra (left
panels) of optimally doped sample with (a) $x=0$ and three overdoped samples
with (b) $x=0$, (c) $x=0.11$, and (d) $x=0.16$. Right panels show the
corresponding optical conductivity spectra. (e) for (a), (f) for (b), (g) for
(c), and (h) for (d), respectively. Solid symbols show the dc conductivity
values at $300$ K. }%
\label{overdoped}%
\end{figure}

With increasing carrier-doping, the characteristics of the optical spectra
change dramatically. Figure \ref{overdoped} shows the reflectivity [(a)-(d)]
and conductivity [(e)-(h)] spectra of the optimally doped sample with $x=0$
and three overdoped samples with $x=0$, $0.11$, and $0.16$, respectively. The
conductivity spectra were calculated from the reflectivity spectra through the
Kramers-Kronig (K-K) transformation. The low frequency conductivity can be
smoothly connected to the dc values determined by dc resistivity measurement,
which confirms the reliability of our K-K transformation. The spectra for the
optimally doped sample with $x=0$ are almost identical to the data published
so far \cite{ZnYBCO2, optimum YBCO1, optimum YBCO2}. The major change with
temperature occurs in the low energy region, as can be seen in Fig.
\ref{detwinn}. In the normal state, the reflectivity increases with lowering
frequency, as expected in a metallic system. In the superconducting state, a
well known step-like behavior \cite{bosonic1, bosonic2, bosonic3} is seen
below $1000$ cm$^{-1}$.%

This behavior is clearly seen in the optimally doped sample while it weakens
with overdoping and is almost suppressed for the heavily overdoped sample.
Moreover, the reflectivity below $200$ cm$^{-1}$ at $10$ K becomes lower with
increasing $p$ and $x$, which implies an increase of some absorption within
the superconducting gap.

In the conductivity spectra, for the Ca-free sample, the $\sigma_{1}%
$-suppression at low frequencies associated with the superconducting
condensation can be seen while the suppression weakens in the overdoped sample
[Fig. \ref{overdoped}(f)]. With further increasing $p$ ($p=0.2$ and $0.22$), a
clear superconducting gap could not be observed, except for a small
suppression of the conductivity, which indicates possible gapless
superconductivity in this state. In this overdoped region, $\sigma_{1}%
(\omega)$ shows a distinct Drude-like increase toward $\omega=0$. The
Drude-like peak for $p=0.22$ is broader than that for $p=0.2$, which indicates
the increase of the residual conductivity with doping, namely, the increase of
the unpaired normal carriers. The calculated spectral weights ($SW=\int%
_{100}^{600}\sigma(\omega)d\omega$) of the residual part at $10$ K are
$4.31\times10^{6}$ $\Omega^{-1}$cm$^{-2}$, $8.01\times10^{6}$ $\Omega^{-1}%
$cm$^{-2}$, $19.92\times10^{6}$ $\Omega^{-1}$cm$^{-2}$, and $30.15\times
10^{6}$ $\Omega^{-1}$cm$^{-2}$ for samples in Figs. \ref{overdoped}(e), (f),
(g), and (h), respectively.

\begin{figure}[ptb]%
\centering
\includegraphics[
height=3.3918in,
width=3.3918in
]%
{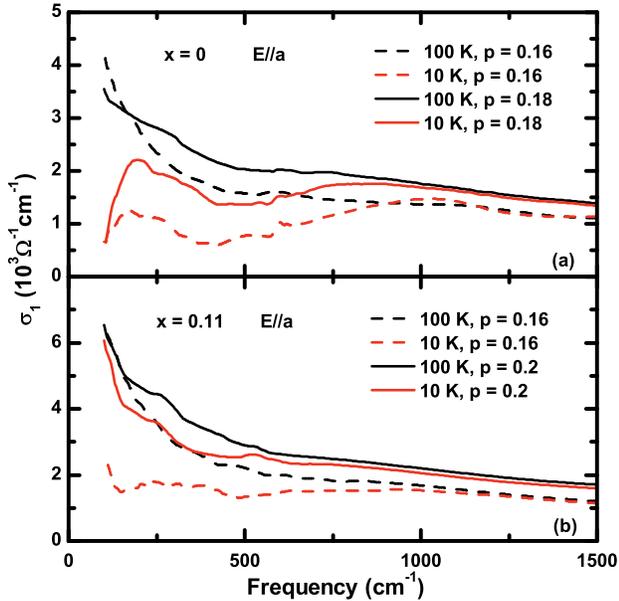}%
\caption{Comparisons of \textit{a}-axis optical conductivity in the normal
state ($100$ K) and the superconducting state ($10$ K) for YBa$_{2}$Cu$_{3}%
$O$_{7-\delta}$ with (a) $x=0$ and (b) $x=0.11$. In both panels, the spectra
of the optimally doped sample ($p=0.16$) and the overdoped sample ($p=0.18$ or
$p=0.2$) are compared. }%
\label{overdoping effect}%
\end{figure}

The strong increase of residual conductivity in Fig. \ref{overdoped} seems to
be the effect of carrier-overdoping. However, it is not obvious whether this
is only due to overdoping or if the disorder introduced by Ca-substitution
contributes. To extract the overdoping effect solely, we compare the spectra
of optimally doped and overdoped samples with the same Ca-content. Figures
\ref{overdoping effect}(a) and (b) show the normal state ($100$ K) and
superconducting state ($10$ K) $\sigma_{1}(\omega)$ in the optimally doped and
in the overdoped region for $x=0$ and $0.11$, respectively. For both
Ca-contents, the missing area is reduced with carrier overdoping. For the
Ca-free sample, the residual conductivity at the lowest temperature ($10$ K)
is enhanced in the overdoped sample, compared to that of the optimally doped
sample. For $x=0.11$, this increase is more pronounced. As a result, the
suppression due to superconducting transition is almost erased by the
Drude-like increase of the conductivity. Since we can observe a similar
increase of residual conductivity for two sets of samples with different
Ca-content, it is concluded that the carrier-overdoping intrinsically enhances
the residual conductivity.%

\subsection{Ca-substitution effect at the optimum doping}

\begin{figure}[ptb]%
\centering
\includegraphics[
height=2.7899in,
width=3.3918in
]%
{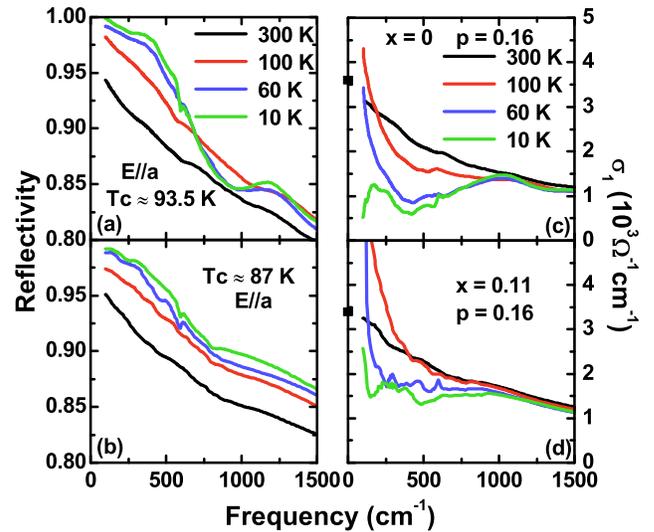}%
\caption{Temperature dependent \textit{a}-axis reflectivity spectra (left
panels) in the optimally doped region for (a) $x=0$ and (b) $x=0.11$. Right
panels show the corresponding optical conductivity spectra. Solid symbols show
the dc conductivity values for $300$ K. }%
\label{Ca-doping effect}%
\end{figure}

Since we work with the Ca-substituted samples, it is necessary to examine
whether the Ca-disorder effect exists. Figure \ref{Ca-doping effect} is a
comparison of the spectra of Ca-free and Ca-substituted samples with the same
doping level ($p=0.16$). In Figs. \ref{Ca-doping effect}(a) and (b) we found
that the step-like feature in the reflectivity spectrum is getting weaker with
Ca-substitution. The reflectivity below $\sim400$ cm$^{-1}$ is slightly lower
for the Ca-substituted sample, which indicates some absorption due to normal carriers.

Figures \ref{Ca-doping effect}(c) and (d) are the corresponding optical
conductivity spectra. Since the doping levels are the same ($p=0.16$) in both
samples, the differences can be attributed solely to the Ca-substitution
effects. A noticeable change with Ca-substitution is seen in the
superconducting state, while there is no substantial difference between the
normal state spectra for $x=0$ and $0.11$. The suppression of the conductivity
due to the superconducting condensation becomes weaker and the missing area
decreases with Ca-substitution. Therefore, we conclude that the increase of
residual conductivity and the decrease in the missing area are also caused by
Ca-substitution, not only overdoping. Here it should be noted that the normal
state spectra for the optimally doped samples do not appreciably change with
Ca-substitution, which is consistent with the dc conductivity behavior.%

\subsection{\textit{b}-axis spectra and the superconducting chain behavior}

\begin{figure}[ptb]%
\centering
\includegraphics[
height=3.9859in,
width=3.3927in
]%
{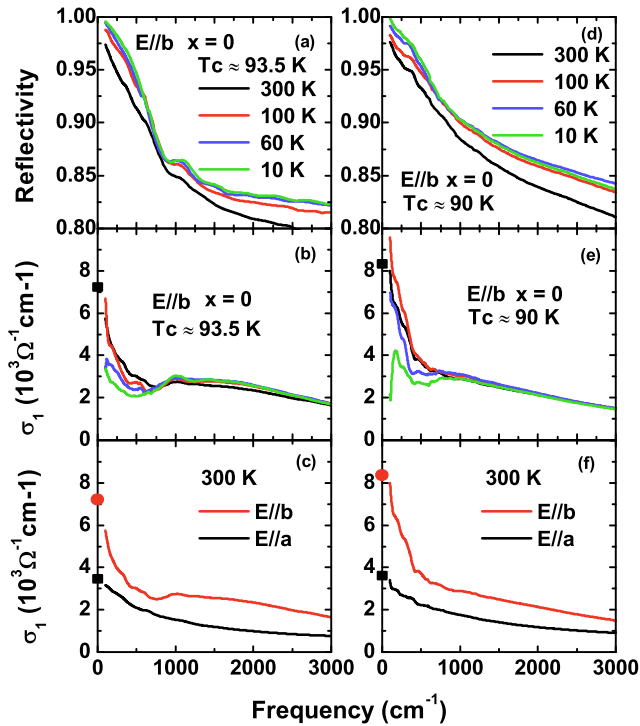}%
\caption{Temperature dependent \textit{b}-axis reflectivity spectra for $x=0$
in the (a) optimally doped region and (d) overdoped region. (b) and (e) show
the corresponding optical conductivity spectra for (a) and (d), respectively.
(c) and (f) show the \textit{a}-axis and \textit{b}-axis optical conductivity
in the room temperature. Solid symbols show the dc conductivity values at
$300$ K. }%
\label{b-axis}%
\end{figure}

So far, we have focused on the \textit{a}-axis spectra which give information
about the two-dimensional CuO$_{2}$ planes. Next, we discuss the
\textit{b}-axis polarized spectra. \textit{b}-axis measurements are also
important for the Y-123 system to extract the chain contribution. Figures
\ref{b-axis}(a) and (d) show the low-frequency \textit{b}-axis reflectivity
spectra for the optimally doped and overdoped YBa$_{2}$Cu$_{3}$O$_{7-\delta}$,
respectively, while Figs. \ref{b-axis}(b) and (e) are the corresponding
optical conductivity. The metallic behavior in the normal state and the
suppression of the conductivity with the superconducting transition can be
seen as expected. The mid-infrared absorption due to disordered chains
\cite{disordered chains1, disordered chains2} between $\sim1000$ cm$^{-1}$ and
$\sim3000$ cm$^{-1}$ is shifting towards lower frequencies with overdoping.
The \textit{b}-axis conductivity values are higher than the \textit{a}-axis
values (Figs. \ref{b-axis}(c) and (f)), which indicates the chain contribution
to the optical conductivity.%

\begin{figure}[ptb]%
\centering
\includegraphics[
height=3.3918in,
width=3.3918in
]%
{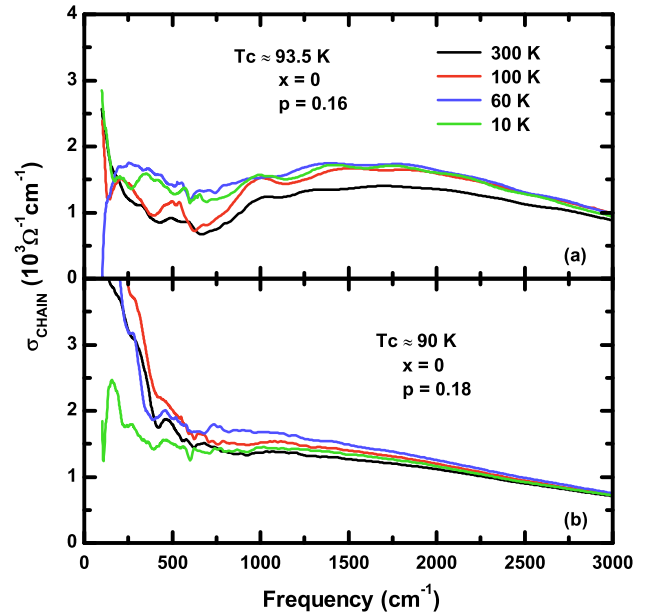}%
\caption{Temperature dependent chain conductivity for $x=0$ in the (a)
optimally doped region and (b) overdoped region.}%
\label{chain}%
\end{figure}

The optical conductivity for the chain ($\sigma_{CHAIN}$) is estimated by
subtracting the \textit{a}-axis conductivity ($\sigma_{1a}$) from the
\textit{b}-axis one ($\sigma_{1b}$) as $\sigma_{CHAIN}=\sigma_{1b}-\sigma
_{1a}$. Figures \ref{chain}(a) and (b) show the chain conductivity in the
optimally doped and overdoped sample, respectively. The conductivity in the
disordered chains will occur with the hopping mechanism \cite{hopping}. On the
other hand, for the highly ordered chain structure, a metallic behavior can be
expected. As support for this idea, we can see the shift of the mid-infrared
absorption to lower energies and the development of a Drude-like spectrum in
the chain response of the highly oxygenated sample. Moreover, at $10$ K, a
distinct suppression due to superconducting condensation can be observed. So
far, although the higher conductivity in the \textit{b}-axis spectra has been
considered to indicate the contribution of the chains to the electronic
conductivity, most of the reported chain spectra showed neither a Drude-like
profile nor a superconducting response. Figure \ref{chain} demonstrates that
with decreasing oxygen deficiency, the chains contribute to the superfluid
density, as well. The superfluid response of the intrinsically metallic CuO
chains would be the result of the proximity effect between chains and
superconducting CuO$_{2}$ planes, which is theoretically predicted in Ref. 20.
Although the missing area in the chain conductivity spectrum was reported for
the oxygen deficient Y-123 in Ref. 21, we observed such a superconducting
response only in the highly oxygenated Y-123.

Another possible source of the anisotropy of $\sigma_{1b}-\sigma_{1a}$ may be
the stripe effect within the CuO$_{2}$ plane. A clear in-plane anisotropy was
observed in the spectra of lightly doped La$_{2-x}$Sr$_{2}$CuO$_{4}$
\cite{underdopoed LSCO} and YBa$_{2}$Cu$_{3}$O$_{y}$ \cite{underdoped YBCO},
which was attributed to the nematic nature of the charge stripes. In Y-123,
the CuO chain is supposed to play a role in alignment of the stripes, while it
has almost no direct contribution to the \ conductivity. Compared to this data
for the underdoped Y-123 \cite{underdoped YBCO}, the magnitude of the
difference $\sigma_{1b}-\sigma_{1a}$ is very large in the present result,
which indicates a huge contribution of the CuO chain to the \textit{b}-axis
conductivity. Therefore, we believe that the anisotropy observed in this study
is mainly caused by the CuO chain conductivity and the effect of the stripes,
even if it exists, is much weaker than the chain conductivity.%

\section{Discussions}

\begin{figure}[ptb]%
\centering
\includegraphics[
height=2.3964in,
width=3.3918in
]%
{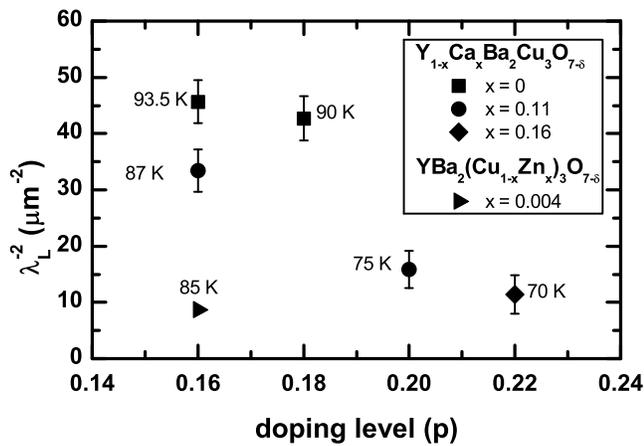}%
\caption{The doping dependence of the superfluid density ($\sim\lambda
_{L}^{-2}$) of Y$_{1-x}$Ca$_{x}$Ba$_{2}$Cu$_{3}$O$_{7-\delta}$ single crystals
for $x=0$, $0.11$ and $0.16$. The $\blacktriangleright$ shows the superfluid
density of YBa$_{2}$(Cu$_{1-x}$Zn$_{x}$)$_{3}$O$_{7-\delta}$ for comparison
(Ref. 2). Temperatures are \textit{T}$_{c}$ values for each sample.}%
\label{missing area}%
\end{figure}

It is possible to calculate the superfluid density with two different methods.
The first method is through the missing area of the \textit{a}-axis optical
conductivity. We calculated the missing area, $A=\int_{100}^{10000}%
[\sigma_{1n}(\omega)-\sigma_{1s}(\omega)]d\omega=c^{2}/8\lambda_{L}^{2}=4\pi
n_{s}e^{2}/m^{\ast}$, using only reliable data down to $100$ cm$^{-1}$ for all
the samples. Here, the spectra at $100$ K and $10$\ K were used for
$\sigma_{1n}(\omega)$ and $\sigma_{1s}(\omega)$, respectively. The obtained
values for $\lambda_{L}^{-2}(\sim$ $\omega_{ps}^{2})$ are plotted in Fig.
\ref{missing area} as a function of the doping level (\textit{p}). The second
method is estimation from the imaginary part of the conductivity, $\sigma
_{2}(\omega)$, which is equal to $\omega_{ps}^{2}/4\pi\omega$ at $\omega
\ll2\Delta$. The values obtained by the two methods are in good agreement with
each other within the error bars in Fig. \ref{missing area}. The $\lambda_{L}$
value for the optimally doped Ca-free sample was determined as $1520$ \AA .
This value is consistent with the reported data \cite{ZnYBCO2, reported pd}.
In Fig. \ref{missing area}, we see a systematic decrease in superfluid density
($n_{s}/m^{\ast}$) with increasing p as well as with increasing
Ca-concentration. Although there is an effect from Ca-substitution, as we
showed in previous section, it is demonstrated that carrier overdoping is also
an effective factor. Distinguishing the Ca-disorder effect from the overdoping
effect, we can conclude that unpaired carriers are intrinsically induced by
overdoping. The decrease of $\lambda_{L}^{-2}$ with overdoping was also
observed in muon spin rotation ($\mu SR$) measurements for Tl2201 \cite{musr1,
musr2} and Ca-doped Y-123 \cite{musr3}, which suggests an electronic
inhomogeneity in these systems.%

Scanning tunnelling microscopy (STM) studies revealed nanoscale spatial
variations in the electronic state of the overdoped Bi$_{2}$Sr$_{2}$Cu$_{2}%
$O$_{y}$ \cite{STM1, STM2}, contrary to many models of high temperature
superconductors that consider electronically homogeneous systems. The increase
of normal carriers in the superconducting state in the overdoped regime with
increasing carrier density is also suggested by other methods. The loss of the
specific heat anomaly and the increase in the zero-temperature specific heat
coefficient [$\gamma(0)$] for (La,Sr)$_{2}$CuO$_{4}$ and TlSr$_{2}$CaCu$_{2}%
$O$_{7-\delta}$ indicate the strong increase of unpaired carriers in the
overdoped region \cite{specific heat}. In NMR measurements, the T-linear
dependence of the nuclear spin-lattice relaxation rate (1/T$_{1}$) was
observed at low temperatures in overdoped TlSr$_{2}$CaCu$_{2}$O$_{7-\delta}$,
which was interpreted to reflect the existence of the residual density of
states at the Fermi level, suggesting gapless superconductivity \cite{NMR}.
Moreover, recently the decrease of the Meissner volume fraction was observed
on field cooling of the magnetic susceptibility measurements \cite{meissner}.
This indicates a decrease in superconducting carrier density due to the phase
separation in the overdoped (La,Sr)$_{2}$CuO$_{4}$.

\begin{figure}[ptb]%
\centering
\includegraphics[
height=7.1996in,
width=3.4039in
]%
{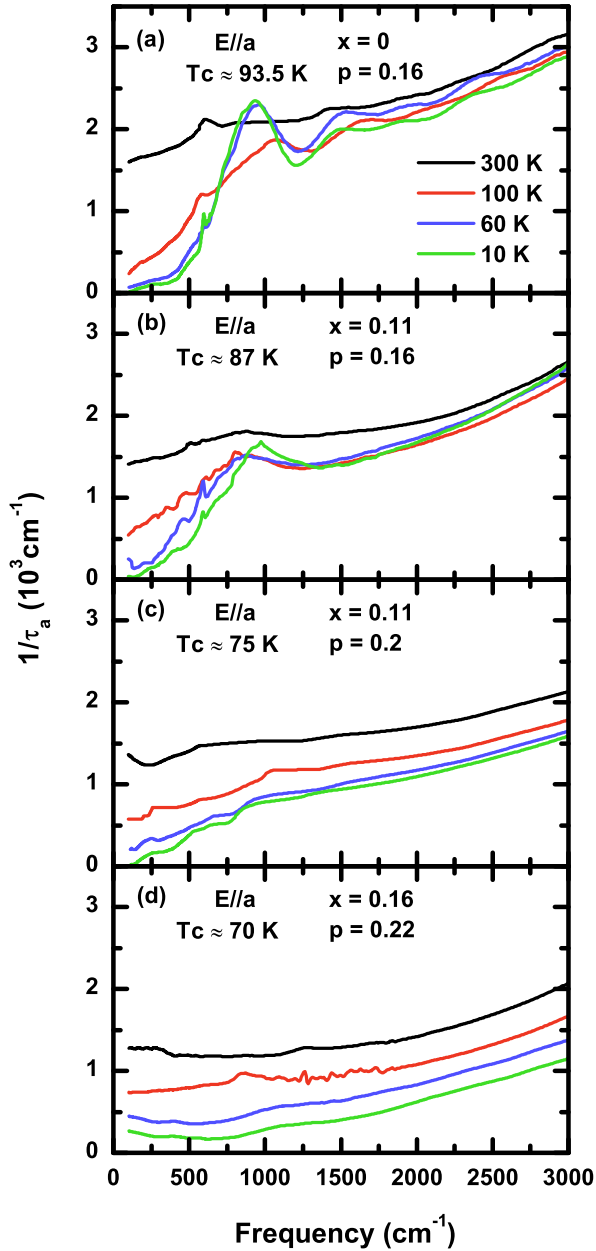}%
\caption{Temperature dependent a-axis optical scattering rate in the optimally
doped region for (a) $x=0$ and (b) $x=0.11$ and in the overdoped region for
(c) $x=0.11$, and (d) $x=0.16$.}%
\label{scattering rate}%
\end{figure}

Finally, we discuss the Ca-substitution effects on the scattering rate. The
\textit{a}-axis scattering rate was calculated by using the equation,
$1/\tau(\omega)=\omega_{p}^{2}/4\pi\cdot\operatorname{Re}(1/\sigma(\omega))$,
as shown in Fig. \ref{scattering rate}. The $\omega_{p}$ values are determined
with the Ferrell-Glover-Tinkham (FGT) sum rule by integration up to
$\omega=10000$ cm$^{-1}$ as $\sim1.66\times10^{4}$ cm$^{-1}$, $\sim
1.69\times10^{4}$ cm$^{-1}$, $\sim1.92\times10^{4}$ cm$^{-1}$, and
$\sim2.29\times10^{4}$ cm$^{-1}$ for $x=0$, $x=0.11$ (optimally doped),
$x=0.11$ (overdoped), and $x=0.16$, respectively. For the optimally doped
sample with x = 0 (Fig. \ref{scattering rate}(a)), the normal state scattering
rates show\ an almost linear increase with $\omega$ up to $\sim3000$ cm$^{-1}%
$. At the superconducting transition, a clear bump is formed around $\sim1000$
cm$^{-1}$ and the low $\omega$ scattering rate is completely suppressed. This
superconductivity-induced feature is getting weaker with Ca-substitution, as
is expected from Fig. \ref{Ca-doping effect}.

When the doping level is further increased, several changes are observed.
First, the bump around $\sim1000$ cm$^{-1}$ vanishes gradually with the
increasing doping level and the \textit{T}-independent high frequency part
becomes \textit{T}-dependent (Figs. \ref{scattering rate}(c) and (d)). The
disappearance of the bump feature is not only due to Ca-substitution but also
due to overdoping (comparison of Figs. \ref{scattering rate}(b) and (c)). The
second noticeable change is the decrease in scattering rate in spite of the
increase of Ca-disorder. It is considered that the carrier doping effect
overcomes the effect of disorder. Thirdly, the $\omega$-dependence changes
from $\omega$-linear towards $\omega^{2}$- behavior. The latter is an
indication of the Fermi liquid nature of the system. For the heavily overdoped
region a small upturn is also observed below $500$ cm$^{-1}$ in Fig.
\ref{scattering rate}(d), which becomes stronger with decreasing temperature.
A similar behavior observed in the heavily overdoped region is discussed in
detail for Bi2212 \cite{sr Bi} and Tl2201 \cite{rcTL2201} systems. The upturn
of the low energy scattering rate usually explained by the charge localization
induced by impurity in disorder introduced systems or by local disordered
octahedral distortion for the Tl2201 system.%

The other important finding in Fig. \ref{scattering rate} is that the
scattering rate is insensitive to Ca-concentration. It is seen when we compare
the spectra for the same doping level (See Figs. \ref{scattering rate}(a) and
(b)). The Ca-insensitive behavior was also observed in the conductivity
spectra and the dc conductivity values in Fig. \ref{Ca-doping effect}. These
facts imply that the Ca atoms in the Y-layer do not act as strong scattering
centers for the carriers in the CuO$_{2}$ planes. The effect of disorders in
the blocking layers was intensively discussed in terms of a weak scattering
center in ref. 34. The Ca-substitution in the present system is considered to
give a similar weak scattering effect.

This weak scattering is in sharp contrast to the strong scattering due to the
in-plane disorder such as in Zn. Zn is a well-known impurity which causes
pair-breaking in a \textit{d}-wave superconductor. The strong Zn-effect can be
seen in the radical reduction of superfluid density in Fig. \ref{missing area}%
, where the $\lambda_{L}^{-2}$ estimated from the optical spectrum of
Zn-substituted Y-123 (Y-123(Zn)) is also plotted \cite{ZnYBCO2}. Although
\textit{T}$_{c}$ values are almost the same (87K and 85K) in optimally doped
Y/Ca-123 and Y-123(Zn), the superfluid density is much smaller in Y-123(Zn)
than in Y/Ca-123. Therefore, in the case of Ca-substitution, the
\textit{T}$_{c}$-suppression accompanied with the reduction in superfluid
density and with the increase in unpaired carriers must be distinguished from
the usual impurity pair-breaking in a \textit{d}-wave superconductor. We need
another mechanism to explain the \textit{T}$_{c}$ suppression without
increasing residual resistivity. It should give the answer for the problem of
the blocking layer effect on \textit{T}$_{c}$.

\section{Conclusions}

We performed temperature dependent measurements of in-plane reflectivity on
detwinned Y$_{1-x}$Ca$_{x}$B$_{2}$Cu$_{3}$O$_{7-\delta}$ single crystals with
various Ca and oxygen concentrations. In the heavily overdoped region, despite
a small suppression of the optical conductivity, a clear superconducting gap
feature could not be observed because of the huge residual conductivity. The
comparison of the spectra between optimally doped and heavily overdoped
samples with the same Ca-contents showed that the residual conductivity
increases with carrier-overdoping. Since we successfully removed the
Ca-disorder effect in the experiment, we conclude that the increase of
residual conductivity is an intrinsic property of the high temperature cuprate superconductors.

We also compared the Ca-substituted and Ca-free samples at the fixed doping
level ($p=0.16$) to observe the Ca-substitution effect on the residual
conductivity. As a result, we found that the disorder effect due to
Ca-substitution also causes the increase of the residual conductivity.

Most of the \textit{b}-axis spectra are characterized by the mid-infrared
absorption due to hopping conduction in the chain in addition to the plane
response. Only for the heavily oxygenated Ca-free Y-123 sample, where the
one-dimensional chain structure shows a highly ordered structure, the chain
itself shows a suppression of conductivity below \textit{T}$_{c}$. A possible
reason for the superconducting chain behavior would be the proximity coupling
effect of the chains with two-dimensional superconducting CuO$_{2}$ planes.

\section{Acknowledgements}

This work was supported by the New Energy and Industrial Technology
Development Organization (NEDO), and the Grant-in-Aid for Scientific Research
from the Ministry of Education, Culture, Sports, Science and Technology of Japan.

\bigskip

\begin{thebibliography}{99}                                                                                               %


\bibitem {pseudogap state}A. V. Puchkov, D. N. Basov, and T. Timusk, J. Phys.:
Condens. Matter 8, 10049 (1996)

\bibitem {ZnYBCO2}N. L. Wang, S. Tajima, A. I. Rykov, and K. Tomimoto, Phys.
Rev. B 57, R11081 (1998)

\bibitem {RC in all HTSC}D. B. Tanner and T. Timusk, Physical Properties of
High Temperature Superconductors III, edited by D. M. Ginsberg (World
Scientific, Singapore, 1992), p. 363

\bibitem {rcYBCO1}J. Sch\"{u}tzmann, S. Tajima, S. Miyamoto, and S. Tanaka,
Phys. Rev. Lett. 73, 174 (1994)

\bibitem {rcYBCO2}J. Sch\"{u}tzmann, S. Tajima, S. Miyamoto, Y. Sato, and I.
Terasaki, Solid State Comm. 94, 293 (1995)

\bibitem {rcCa-YBCO}C. Bernhard, R. Henn, A. Wittlin, M. Kl\"{a}ser, Th. Wolf,
G. M\"{u}ller-Vogt, C. T. Lin, and M. Cardona, Phys. Rev. Lett. 80, 1762 (1998)

\bibitem {rcTL2201}Y. C. Ma and N. L. Wang, Phys. Rev. B 73, 144503 (2006)

\bibitem {rcBi2212}J. Hwang, T. Timusk, and G. D. Gu, J. Phys.: Condens.
Matter 19, 125208 (2007)

\bibitem {pulling technique}Y. Yamada and Y. Shiohara, Physica C 217, 182 (1993)

\bibitem {p-Tc}M. R. Presland, J. L. Tallon, R. G. Buckley, R. S. Liu, and N.
E. Flower, Physica C 176, 95 (1991)

\bibitem {Fisher}B. Fisher, J. Genossar, C. G. Kuper, L. Patlagan, G. M.
Reisner, and A. Knizhnik, Phys. Rev. B 47, 6054 (1993)

\bibitem {optimum YBCO1}L. D. Rotter, Z. Schlesinger, R. T. Collins, F.
Holtzberg, C. Field, U. W. Welp, G. W. Crabtree, J. Z. Liu, Y. Fang, K. G.
Vandervoort, and S. Fleshlert, Phys. Rev. Lett. 67, 2741 (1991)

\bibitem {optimum YBCO2}D. N. Basov, R. Liang, B. Dabrowski, D. A. Bonn, W. N.
Hardy, and T. Timusk, Phys. Rev. Lett., 77, 4090 (1996)

\bibitem {bosonic1}D. N. Basov and T. Timusk, Rev. Mod. Phys. 77, 721 (2005)

\bibitem {bosonic2}J. J. Tu, C. C. Homes, G. D. Gu, D. N. Basov, and M.
Strongin, Phys. Rev. B 66, 144514 (2002)

\bibitem {bosonic3}N. L. Wang, P. Zheng, J. L. Luo, Z. J. Chen, S. L. Yan, L.
Fang, and Y. C. Ma, Phys. Rev. B 68, 054516 (2003)

\bibitem {disordered chains1}D. N. Basov, R. Liang, D. A. Bonn, W. N. Hardy,
B. Dabrowski, M. Quijada, D. B. Tanner, J. P. Rice, D. M. Ginsberg, and T.
Timusk, Phys. Rev. Lett. 74, 598 (1995)

\bibitem {disordered chains2}Z. Schlesinger, R. T. Collins, F. Holtzberg, C.
Feild, S. H. Blanton, U. Welp, G. W. Crabtree, Y. Fang, and J. Z. Liu Phys.
Rev. Lett. 65, 801 (1990)

\bibitem {hopping}J. Sch\"{u}tzmann, B. Gorshunov, K.\ F. Renk, J. M\"{u}nzel,
A. Zibold, H. P. Geserich, A. Erb, and G. M\"{u}ller-Vogt, Phys. Rev. B 46,
512 (1992)

\bibitem {theoretical superconducting chain}V. Z. Kresin and S. A. Wolf, Phys.
Rev. B 46, 6458 (1992)

\bibitem {proximity1}Y. -S. Lee, K. Segawa, Y. Ando, and D. N. Basov, Phys.
Rev. Lett. 94, 137004 (2005)

\bibitem {underdopoed LSCO}M. Dumm, S. komiya, Y. Ando, and D. N. Basov, Phys.
Rev. Lett. 91, 077004 (2003)

\bibitem {underdoped YBCO}Y. -S. Lee, K. Segawa, Y. Ando, and D. N. Basov,
Phys. Rev. B 70, 014518 (2004)

\bibitem {reported pd}D. N. Basov, R. Liang, D. A. Bonn, W. N. Hardy, B.
Dabrowski, M. Quijada, D. B. Tanner, J. P. Rice, D. M. Ginsberg, and T.
Timusk, Phys. Rev. Lett. 74, 598 (1995)

\bibitem {musr1}Ch. Niedermayer, C. Bernhard, U. Binninger, H. Gl\"{u}kler, J.
L. Tallon, E. J. Ansaldo, and J. I. Budnick, Phys. Rev. Lett. 71, 1764 (1993)

\bibitem {musr2}Y. J. Uemura, A. Keren, L. P. Le, G. M. Luke, W. D. Wu, Y.
Kubo, T. Manako, Y. Shimakawa, M. Subramanian, J. L. Cobbs, and J. T. Markert,
Nature 364, 605 (1993)

\bibitem {musr3}J. L. Tallon, C. Bernhard, U. Binninger, A. Hofer, G. V. M.
Williams, E. J. Ansaldo, J. I. Budnick, and Ch. Niedermayer, Phys. Rev. Lett.
74, 1008 (1995)

\bibitem {STM1}H. Mashima, N. Fukuo, Y. Matsumoto, G. Kinoda, T. Kondo, H.
Ikuta, T. Hitosugi, and T. Hasegawa, Phys. Rev. B 73, 060502R (2006)

\bibitem {STM2}K. Kudo, T. Nishizaki, N. Okumura, and N. Kobayashi, 25th
International Conference on Low Temperature Physics, Journal of Physics:
Conference Series 150, 052133 (2009)

\bibitem {specific heat}J. W. Loram, K. A. Mirza, J. M. Wade, J. R. Cooper,
and W. Y. Liang, Physica C 235, 134 (1994)

\bibitem {NMR}K. Magishi, Y. Kitaoka, G.-q. Zheng, K. Asayama, T. Kondo, Y.
Shimakawa, T. Manako, and Y. Kubo, Phys. Rev. B 54, 10131 (1996)

\bibitem {meissner}Y. Tanabe, T. Adachi, T. Noji, and Y. Koike, J. Phys. Soc.
Jpn. 74, 2893 (2005)

\bibitem {sr Bi}J. Hwang, T. Timusk, and G. D. Gu, Nature 427, 714 (2004)

\bibitem {weak scattering center}K. Fujita, T. Noda, K. M. Kojima, H. Eisaki,
and S. Uchida, Phys. Rev. Lett. 95, 097006 (2005)
\end{thebibliography}
\end{document}